\documentstyle[psfig]{mn}

\title[Optical and X-ray observations of OY Car]{Optical and {\it\bf ROSAT\/} X-ray observations of the dwarf nova OY Carinae in superoutburst and quiescence}

\author[G.W. Pratt et al.]
	{Gabriel W. Pratt,$^1$ B.J.M. Hassall,$^1$ T. Naylor,$^2$ Janet H. Wood$^2$ and J. Patterson$^3$ \\
	 $^1$Centre for Astrophysics, University of Central Lancashire, Preston PR1 2HE, U.K.\\
	 $^2$Department of Physics, Keele University, Keele, ST5 5BG, U.K.\\
	 $^3$Department of Astronomy, Columbia University, 538 West 120th Street, New York, New York 10027, U.S.A.}
         
\begin{document}

\maketitle


\begin{abstract}

We present {\it ROSAT} X-ray and optical light curves of the 1994 February superoutburst of the eclipsing SU UMa dwarf nova OY Carinae. There is no eclipse of the flux in the {\it ROSAT\/} HRI light curve. Contemporaneous `wide $B$' band optical light curves show extensive superhump activity and dips at superhump maximum. Eclipse mapping of these optical light curves reveals a disc with a considerable physical flare, even three days into the superoutburst decline. 

We include a later (1994 July) {\it ROSAT\/} PSPC observation of OY Car that allows us to put constraints on the quiescent X-ray spectrum. We find that while there is little to choose between OY Car and its fellow high inclination systems with regard to the temperature of the emitting gas and the emission measure, we have difficulties reconciling the column density found from our X-ray observation with the column found in {\it HST\/} UV observations by Horne et al. (1994). The obvious option is to invoke time variability.

\end{abstract}




\section{Introduction}

Dwarf novae are a subclass of non-magnetic cataclysmic variable (CV), binary systems consisting of a Roche lobe filling late-type secondary star in the process of mass transfer, via an accretion disc, onto a white dwarf primary. Sudden increases in the accretion rate through the disc and onto the white dwarf, most likely due to thermal instability, cause outbursts of $\sim 2-5$ magnitudes on timescales of weeks to years. The SU UMa subclass undergo additional infrequent superoutbursts, brighter by $\sim 1$ mag and lasting $\sim 5$ times longer than ordinary outbursts, thought to be triggered by a tidal resonance of the material in the outer disc. See Warner (1995) for a comprehensive review. 

Steady state theory predicts that half the accretion luminosity from these systems is released in the disc, the other half in the boundary layer between the disc and the white dwarf. This is the region where the material of the disc is decelerated to match the surface velocity of the primary, and in quiescence, it is believed to consist of an optically thin gas emitting relatively hard ($\sim 2 - 20$ keV) X-rays. In outburst, the accretion rate rises and the X-ray emitting region is predicted to become optically thick, with a characteristic emission of softer ($\sim 0.1 - 1.0$ keV) X-rays (Pringle \& Savonije 1979). Provided the primary is not spinning fast, the boundary layer is thought to be the dominant source of the X-ray emission from non-magnetic CVs \cite{pringle77}.

\begin{figure*}
\centerline{\psfig{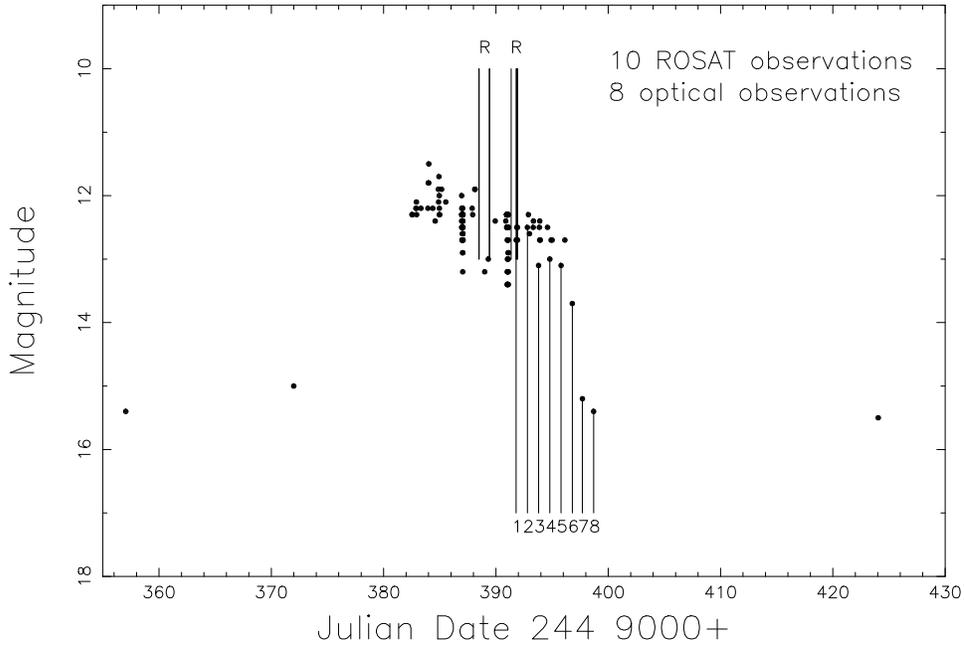}}
\caption{The light curve of OY Car during the 1994 Jan superoutburst, from observations supplied by the VSS RAS NZ. Our optical data, obtained on the decline, and the times of the {\it ROSAT\/} observations (R) are also shown. The optical data have been converted to Oke AB magnitudes using $AB = -2.5\log{f_{\nu}} - 48.60$ (Oke 1974).}\label{fig:contxtlcrv}   
\end{figure*}

Observations have, to a great extent, borne out the theoretical predictions. Those systems where the white dwarf, bright spot and accretion disc are all eclipsed allow in principle the simplest observational test to determine the source of the X-ray emission: if it is in the vicinity of the white dwarf, as expected for a boundary layer, then there will be an eclipse of the X-ray flux. In general these X-ray observations are a few tens of kiloseconds in length and require the binning of the data. The location of the X-ray source can be further pinpointed by comparing the X-ray light curves with optical light curves. 

Of the high inclination dwarf novae, three have so far been examined in this way. A {\it ROSAT} observation of HT Cas in the low state showed an eclipse \cite{woodetal95a}; this system was subsequently observed in quiescence by both {\it ROSAT} and {\it ASCA\/} \cite{mukaietal97}, the latter observation of sufficient quality to allow approximate source dimensions to be extracted from the X-ray data. In addition, van Teeseling \shortcite{vant97} has observed an eclipse of the X-ray flux from Z Cha in quiescence, while Pratt et al. (1999) have confirmed that the X-ray source in OY Car is also eclipsed in quiescence. 

Observations of dwarf novae in outburst or superoutburst are difficult due to their inherent unpredictability and the short timescales (at least in terms of satellite response times) involved. Generally, the hard X-ray emission is seen to drop in many systems during outburst, often accompanied by a surge in the soft X-ray flux. The deepest insights into the outburst state can again be derived from observations of the high inclination systems. Perhaps most interestingly, the {\it EXOSAT} observation of OY Car by Naylor et al. (1988) showed that there was no X-ray eclipse during superoutburst. Contemporaneous observations at other wavelengths (Naylor et al. 1987, 1988) inferred considerable vertical structure in the outer disc, causing material to block the view of the central regions at all orbital phases. This interpretation is supported by the lack of eclipse in the {\it ROSAT\/} PSPC observation of the high inclination novalike variable UX UMa (Wood, Naylor \& Marsh 1995), which is believed to be similar to a dwarf nova in a state of permanent outburst. Additionally, advances in eclipse mapping have recently allowed modelling of outburst discs: {\it HST\/} UV observations of Z Cha at normal outburst peak by Robinson et al. \shortcite{robinsonetal95} reveal a flared disc with an opening full-angle of $\alpha = 16^{\circ}$. Thus it appears that in the case of OY Car in superoutburst, we are seeing the X-ray flux from an extended (or `coronal') source, and not from the boundary layer directly, a result of the obscuration due to a physically thick, vertically extended disc. 

Discrepancies exist between boundary layer theory and observation, however. The most famous (and still unsolved) problem remains the fact that the X-ray luminosity rarely approaches the predicted value, being typically down by an order of magnitude on the optical+UV luminosity from the disc. The ratio is most discrepant for the optically thick cases, where the accretion rate is high, such as in the case of dwarf novae in outburst, or novalike variables (van Teeseling \& Verbunt 1994). One possible explanation is that the difference is due to loss of kinetic energy in the wind that is often observed to come from the inner regions of the accretion disc during high mass transfer episodes. Another is a combination of this wind energy loss in conjunction with enhanced absorption due to the outburst. {\it EUVE} observations of U Gem in outburst have shown that the spectrum is composed of emission lines, likely produced in the corona or wind, and a highly absorbed blackbody-like continuum \cite{longetal96}. Another explanation, applicable to both high and low accretion rate systems, may simply be that the white dwarf is spinning fast. The low $L_{X}/L_{opt+UV}$ ratio is also observed in quiescence (e.g., Belloni et al. 1991); here it may be due to the non-steady state conditions caused by the mass accretion rate onto the primary being lower than the mass flux rate through the outer disc.


\begin{table*}
\begin{minipage}{150mm}
\center
\caption{{\small Journal of X-ray observations.}}
\begin{tabular}{|l|c|c|c|r|r|}
\hline

\multicolumn{1}{|l}{State} & \multicolumn{1}{|c}{Date} & \multicolumn{1}{|c}{HJED start} &\multicolumn{1}{|c|}{Phase covered} & \multicolumn{1}{c|}{Duration} & \multicolumn{1}{|c}{Source count rate}\footnote{This is the time-averaged count rate from the source after background subtraction. The errors quoted are Poissonian} \\ 

\multicolumn{1}{|c}{ } & \multicolumn{1}{|l}{ } & \multicolumn{1}{|c}{ } & \multicolumn{1}{|c|}{} & \multicolumn{1}{c|}{(seconds)} & \multicolumn{1}{|c}{(counts s$^{-1} \times 10^{-2}$)} \\ 


Superoutburst &             & (244 9300+) &              &         & \\
              & 1994 Feb 4  & 88.492 & 0.885--0.916 & 160   & $4.8 \pm 1.7$  \\
              & .           & 88.499 & 0.997--0.224 & 1240  & $3.7 \pm 0.5$  \\
              & 1994 Feb 5  & 89.419 & 0.579--0.600 & 112   & $-0.4 \pm 0.6$  \\
              & .           & 89.427 & 0.709--0.945 & 1288  & $3.0 \pm 0.5$  \\
              &             &             &              &         &     \\

              & 1994 Feb 7  & 91.352 & 0.197--0.433 & 1288  & $2.5 \pm 0.4$  \\
              & 1994 Feb 8  & 91.799 & 0.277--0.599 & 1624  & $2.5 \pm 0.4$  \\
              & .           & 91.870 & 0.405--0.428 & 80    & $3.0 \pm 2.0$  \\
              & .           & 91.879 & 0.549--0.716 & 904   & $2.2 \pm 0.5$  \\
              & .           & 91.891 & 0.745--0.839 & 512   & $1.7 \pm 0.6$  \\
	      & .           & 91.930 & 0.353--0.404 & 280   & $3.9 \pm 1.2$  \\

              &             &             &              &  &               \\
              &             &             &              & Average & $2.8 \pm 0.2$  \\
              &             &             &              &  &                \\

Quiescence    &             & (244 9500+) &              &  &                \\
              & 1994 July 5 & 38.916 & 0.003--0.088 & 464   & $8.0 \pm 1.3$  \\
              & .           & 39.313 & 0.296--0.398 & 560   & $8.0 \pm 1.2$  \\
	      & 1994 July 6 & 40.473 & 0.664--0.670 & 32    & $0.5 \pm 1.3$  \\
	      & 1994 July 7 & 40.508 & 0.229--0.328 & 544   & $8.7 \pm 1.3$  \\

              &             &             &              &  &                \\
              &             &             &              & Average & $8.1 \pm 0.7$  \\
              &             &             &              &  &                \\

\hline
\end{tabular}
\label{tab2:1}

\end{minipage}
\end{table*}

OY Car is one of the best studied of the dwarf novae, mainly because it is a fully eclipsing system with an inclination angle of $83^{\circ}$. But with the benefits of a full eclipse come other difficulties, chief among them the effect of the disc, which is seen nearly edge on in OY Car. We know that the disc becomes physically thick during outburst and superoutburst (see e.g., Naylor et al. 1988, Robinson et al. 1995), but we do not know the extent of the contribution this has to the overall rise in the local absorption during these episodes. The extent of the absorption to OY Car, both interstellar and local, is still very much a mystery. Horne et al. \shortcite{hornetal94} use {\it HST} observations of OY Car in quiescence to deduce the existence of a significant local absorption due to a veiling material, subsequently modelled by a solar abundance LTE gas. Horne et al. dubbed this veiling gas the `iron curtain' after Shore (1992); mainly due to the existence of many blended Fe{\scriptsize II} lines in its spectrum. They derive a column density of $n_{H} \simeq 10^{22}$ cm$^{-2}$, which, as pointed out by Naylor \& la Dous \shortcite{naylad97}, is sufficient to absorb much of the soft X-ray emission from the boundary layer. If the absorption is this high in quiescence, how much more local absorption is added during outburst and superoutburst?

The aim of the observations herein is to address not only the extent of the thickening of the disc during superoutburst and the question of the absorption to the X-ray source, but also the missing boundary layer problem. To this end, we present contemporaneous optical and {\it ROSAT\/} X-ray observations of the eclipsing SU UMa system OY Car in superoutburst, together with a later {\it ROSAT\/} X-ray observation of the system in quiescence. We compare past and present X-ray observations of OY Car with previous observations of the canonical low inclination SU UMa system, VW Hydri, in order to illustrate the effects of the disc in the high inclination case. The high time resolution optical photometry, obtained at the end of the superoutburst peak and on the decline, enables us to use a new version of the eclipse mapping method (originally developed by Horne (1985), since updated to include disc flaring, e.g. Robinson et al. 1995, Wood 1994) to investigate the extent of the physical flaring of the disc as the system returns to quiescence. We use the quiescent {\it ROSAT\/} X-ray observation to delimit the extent of the absorption in quiescence, and compare this to the results from the {\it HST} data from Horne et al. \shortcite{hornetal94}. 

Section~\ref{sec:obs_dat} describes the X-ray and EUV observations and reduction, in Section~\ref{sec:xray} we describe the analysis of the X-ray data, Section~\ref{sec:sob_opt} describes the superoutburst optical data, and in Section~\ref{sec:res_disc}, we discuss our results and compare and contrast to those found previously for VW Hyi.


\section{Observations and data reduction}
 \label{sec:obs_dat}

Monitoring observations made by the Variable Star Section of the Royal Astronomical Society of New Zealand (VSS RAS NZ) reveal that OY Car underwent a superoutburst on 1994 February 1, stayed at maximum for 7 days, and returned to quiescence over a subsequent 13 day period. The first X-ray observation, obtained with the {\it ROSAT\/} High Resolution Imager (HRI; David et al. 1991) was taken at the height of the superoutburst. The optical observations were taken at the end of the superoutburst plateau and on the decline, and so the full dataset includes one contemporaneous optical/X-ray observation (1994 Feb 8). The second X-ray observation took place in 1994 July, when the quiescent OY Car was observed with the {\it ROSAT\/} Position Sensitive Proportional Counter (PSPC; Pfeffermann \& Briel 1986). Both the superoutburst and the quiescent {\it ROSAT} observations had simultaneous Wide Field Camera (WFC; Sims et al. 1990) coverage.

Figure~\ref{fig:contxtlcrv} shows the magnitude of OY Car before, during, and after the 1994 Jan superoutburst, using both data from the VSS RAS NZ and our own optical observations. The figure also shows the timing of the X-ray observations relative to the optical observations. Exact times are given in Table~\ref{tab2:1}. 

The orbital period of OY Car is $\sim 91$ min, very similar to the $\sim 96$ min orbit of {\it ROSAT}. The orbital phase of OY Car as seen by the satellite thus precesses through the satellite orbit (with a stepsize of $\sim 5$ min per orbit), necessitating careful scheduling of observations to obtain the required phase coverage.  For the remainder of this paper, to avoid confusion when referring to the X-ray data, we will use `observation' to describe the actual $n$-day baseline over which the individual `datasets' were obtained.


\subsection{X-ray and EUV observations}

\subsubsection{Superoutburst}
\label{sec:sob_redux}


OY Car was observed by the {\it ROSAT\/} HRI for a total of 7.4 kiloseconds between 1994 February 4 and 1994 February 8, with an observation start time of 1994 February 4 23:43:33 {\scriptsize UT} (HJED 244 9388.489 639). The dates, times and durations of the individual X-ray observation datasets are given in Table~\ref{tab2:1}, the journal of X-ray observations. 


The data were reduced using {\scriptsize ASTERIX} software. We extracted source counts from a circular region of radius 150 arcsec. The background annulus, centred on the source, had inner and outer radii of 200 arcsec\ and 600 arcsec\ respectively. These dimensions were chosen for compatibility with those used in the comprehensive compilation of {\it ROSAT} CV observations by van Teeseling, Beuermann \& Verbunt (1996). Source and background data were then binned into 1s bins and background subtracted. The resulting background-subtracted total flux light curve was corrected for vignetting, dead time, and scattering. Lastly, all times were converted to HJED.


The {\it ROSAT} WFC observed OY Car for 26.6 kiloseconds on 1994 February 5 and 5.9 kiloseconds on 1994 February 7, with the S1a filter, which is sensitive to the spectral range 90 - 206 eV. The data were extracted using {\scriptsize ASTERIX} and {\scriptsize WFCPACK} software. Using a source radius of 5\arcmin\ and a background annulus, centred on the source, of inner and outer radii 10\arcmin\ and 20\arcmin\, respectively, we estimate a $3\sigma$ upper limit to the superoutburst count rate of $1.2 \times 10^{-3}$ counts s$^{-1}$.
 

\subsubsection{Quiescence}


OY Car was observed in quiescence with the {\it ROSAT\/} PSPC between 1994 July 5 and 1994 July 7 for a total of 1.6 kiloseconds. OY Car was in the middle of a quiescent spell - the previous normal outburst took place on 1994 May 30, and the subsequent superoutburst occurred on 1994 December 11. The start time of the observations was 1994 July 5 09:55:55 {\scriptsize UT} (HJED 244 9538.915 388). Information about these X-ray observations can also be found in Table~\ref{tab2:1}.


The observations were taken off-axis and with a spacecraft wobble to avoid vignetting by PSPC support wires, so the source and background counts were extracted from larger regions than those used for the superoutburst data. The data were again reduced using the {\scriptsize ASTERIX} package. The radius of the source region used was 250 arcsec, while the inner and outer radii for the background annulus, centred on the source, were 300 arcsec and 900 arcsec, respectively. These dimensions were again chosen for compatibility with van Teeseling et al. (1996). These data were then corrected as described for the superoutburst data. 

In addition, the spectrum from channels 11 to 235 was extracted in 1 second bins. It was then binned, using {\scriptsize FTOOLS}, into 11 spectral bins so that there were at least 5 counts in each bin.


The WFC also observed OY Car for a total of 4.9 kiloseconds on 1994 July 5 with the S1a filter. The data were reduced using {\scriptsize ASTERIX} and {\scriptsize WFCPACK} software. There was again no detection, and using the method described for the superoutburst data, the 3$\sigma$ upper limit found for the quiescent count rate is $1.5 \times 10^{-3}$ counts s$^{-1}$.


\subsection{Optical observations}

High time resolution optical photometry was obtained on the 1m telescope at CTIO, Chile. The dates and times of each observation are presented in Table~\ref{tab2:2:1}, the journal of optical observations. The Automated Single-Channel Aperture Photometer (ASCAP) was used in conjunction with a Hamamatsu R943-02 photomultiplier tube. All observations were taken using an integration time of 5 sec through a CuSO$_{4}$ filter. This filter has a wide bandpass (3300-5700 \AA) which gives a broad $B$ response, but includes a substantial amount of $U$. The counts per 5 sec were corrected for atmospheric extinction and phased with the orbital period using the updated ephemeris of Pratt et al. \shortcite{prattetal99}.

\begin{table}
\center
\caption{{\small Journal of optical observations}}
\begin{tabular}{|l|l|r|}
\hline

\multicolumn{1}{|l}{UT Date} & \multicolumn{1}{|l}{UT start} &\multicolumn{1}{|l|}{Duration} \\ 

\multicolumn{1}{|l}{ } & \multicolumn{1}{|c}{ } & \multicolumn{1}{|l|}{(sec)} \\

\hline

 1994 Feb 8  & 06:21:22 & 9452 \\
 1994 Feb 9  & 07:26:04 & 5647 \\
 1994 Feb 10 & 08:12:09 & 3606 \\
 1994 Feb 11 & 07:24:47 & 6358 \\
 1994 Feb 12 & 07:26:03 & 6324 \\
 1994 Feb 13 & 07:22:58 & 6521 \\
 1994 Feb 14 & 06:32:42 & 8641 \\
 1994 Feb 15 & 06:48:22 & 8427 \\

\hline
\end{tabular}
\label{tab2:2:1}
\end{table}


\section{The X-ray data}
\label{sec:xray}


\subsection{The Superoutburst X-ray light curve}
\label{sec:sob_lcrv}

\begin{figure}
\centerline{\psfig{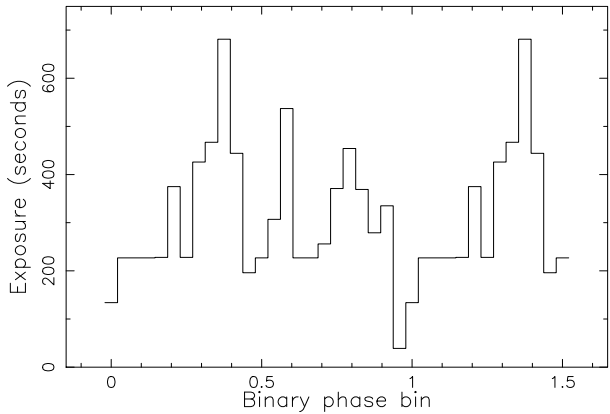}}
\caption{The effective exposure times of each of the 24 phase bins used in the generation of the superoutburst X-ray light curve.}\label{fig:bincoverage}   
\end{figure}

The average count rates decline during the course of the observation, but recover by the last dataset. We divided the observation into two groups along the lines of the natural break in the datasets (Feb 5/Feb 7), and found that there is a negligible drop in the average count rate between the first three datasets, Set 1, and the last five, Set 2. 

\begin{figure*}
\centerline{\psfig{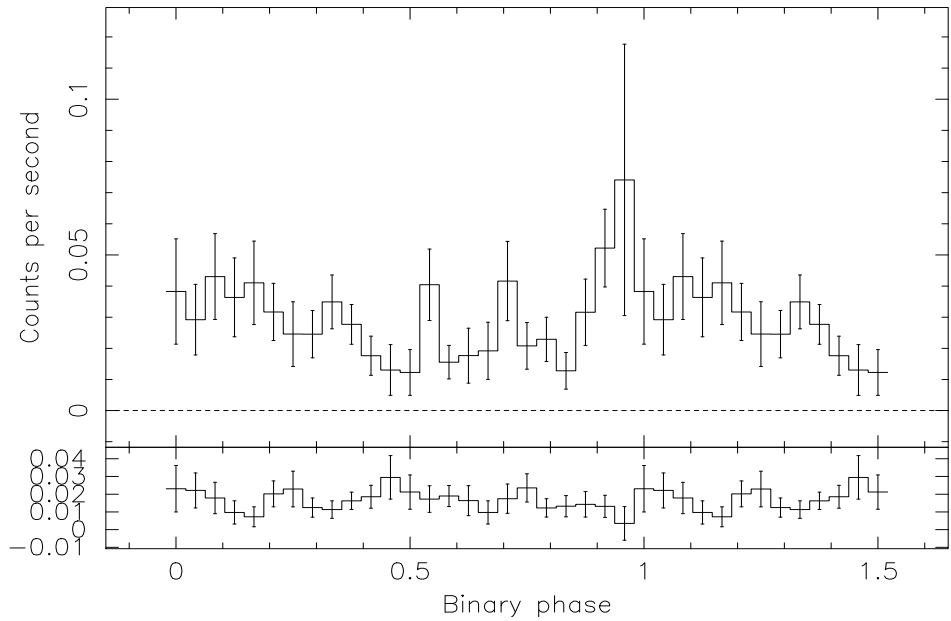}}
\caption{The upper panel shows the {\it ROSAT\/} HRI superoutburst X-ray light curve folded, in 24 bins, on the orbital period of OY Car. The lower panel shows the background light curve scaled to the dimensions of the source extraction circle. All error bars are Poissonian.}\label{fig:soblcrv}
\end{figure*}

The total superoutburst light curve was obtained by folding all the datasets into 24 bins on the orbital period of OY Car using the ephemeris of Pratt et al. \shortcite{prattetal99}. The duration of each bin was marginally smaller, at 227.2 s, than the totality of the white dwarf eclipse as given in Wood et al. (1989), of 231.2 s. The starting point of the binning was chosen so that one phase bin was exactly centred on phase zero. The effective exposure time of each of the 24 bins is illustrated in Figure~\ref{fig:bincoverage}. The phase coverage in the data was such that $\sim 97$ per cent of the orbit was covered, including one eclipse. The resulting phase-folded superoutburst light curve can be seen in Figure~\ref{fig:soblcrv}. The large error bars on the bin centred on phase $\phi = 0.95$ are due to the poor coverage in this bin. This lack of coverage extends $\sim 25$ per cent into the white dwarf eclipse phase bin at phase $\phi = 0.0$. 

The mean out of eclipse flux is $(2.8 \pm 0.2) \times 10^{-2}$ counts s$^{-1}$, while the flux in WD eclipse is $(3.8 \pm 1.7) \times 10^{-2}$ counts s$^{-1}$. We can thus ascertain that there is no total eclipse of the X-rays in the vicinity of the white dwarf at the $2.2\sigma$ level.

If we instead take only the first four datasets, which cover the eclipse phase, and apply the above analysis, we find that there is no eclipse of the X-ray flux at the $2.3\sigma$ level.


\subsection{The Superoutburst Luminosity}

\label{sec:sob_lum}

As the {\it ROSAT\/} HRI data contains no spectral information, we have used the total time-averaged count rate in order to make a rough estimate of the intrinsic luminosity of the system in superoutburst. 

The derivation of an approximate luminosity in superoutburst involved generating simulated spectra within {\scriptsize XSPEC}. As a basis for a temperature estimate, we used $kT = 6.0$ keV, the dominant temperature found for the high energy spectrum of VW Hyi by Wheatley et al. (1996). The simulated spectra were normalised to the count rate of our observation, with intervening absorptions of $10^{20}, 10^{21}$, and $10^{22}$ cm$^{-2}$ included in the models. The $n_{H}$ parameter is completely unknown for the superoutburst case. As discussed in the introduction, it is likely that the {\it local} absorption becomes the dominant contributor to the global column in superoutburst as a result of the inclination of the system. Removal of the estimated total intervening absorption allowed the calculation of the corresponding intrinsic luminosities. Using a distance of 82 pc (Wood et al. 1989) and a bandpass of 0.001 to 20 keV for a `bolometric' value, the resulting superoutburst luminosity as deduced by this method is $L_{X} \sim 2 \times 10^{32}$ erg s$^{-1}$.


\subsection{Spectral analysis of the quiescent data}


\begin{figure*}
\begin{minipage}{150mm}
\centerline{\psfig{figure=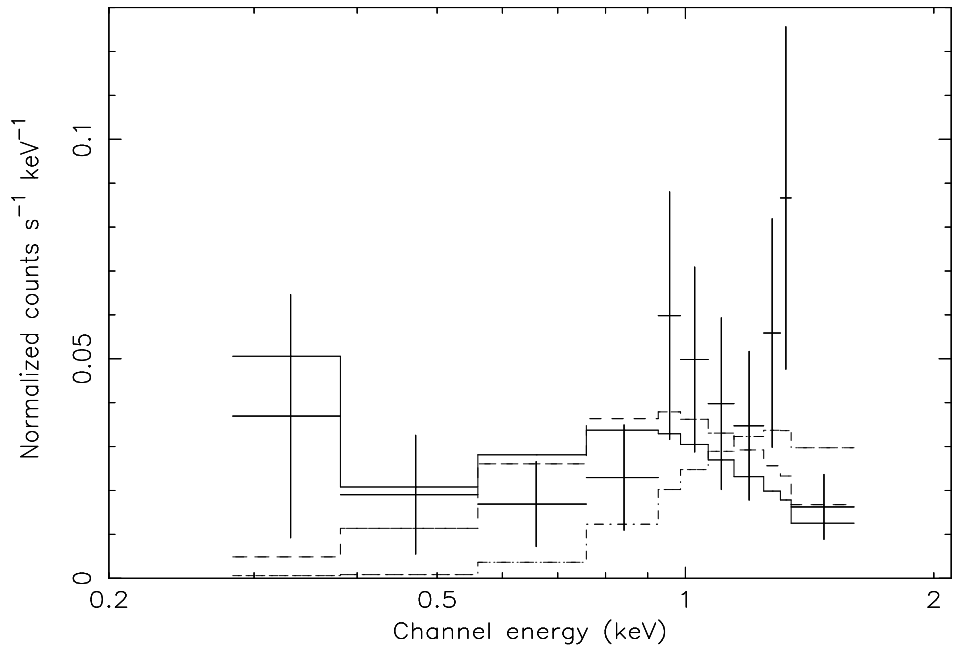,height=9.5cm,width=14cm,angle=270}}
\caption{The {\it ROSAT\/} PSPC spectrum of OY Car in quiescence. Each line shows a thermal Bremsstrahlung  spectrum with $kT = 2.4$ keV folded through the instrumental response. The different absorbing columns are: solid: $n_{H} = 10^{20}$ cm$^{-2}$, long dash: $n_{H} = 10^{21}$ cm$^{-2}$, and dot-dash: $n_{H} = 10^{22}$ cm$^{-2}$.}\label{fig:quiespec}
\end{minipage}
\end{figure*}

The PSPC data obtained in quiescence covered only $\sim 30 $ per cent of the orbital phase with no coverage at phase zero, making it impossible to search for an X-ray eclipse of the boundary layer. We were, however, able to use the spectral information of the PSPC observations to place constraints on the X-ray emitting source at quiescence.

First, the PSPC spectrum was binned into the photon channel bandpasses summarised in Table~\ref{tab3:3:1}, and the count rates for each bandpass calculated. The average hardness ratio is defined as $(B-A)/(B+A)$; our analysis indicates a value of $(B-A)/(B+A) = 0.77 \pm 0.22$ for OY Car in quiescence.

\begin{table}
\center
\caption{{\small Photon-channel bandpasses for the binned quiescent spectrum of OY Car}}
\begin{tabular}{|l|r|c|r|}
\hline

\multicolumn{1}{|c}{Bandpass} & \multicolumn{1}{|c|}{Channels} &\multicolumn{1}{c|}{Energy} & \multicolumn{1}{c|}{Count Rate} \\ 

\multicolumn{1}{|c}{ } & \multicolumn{1}{|c|}{ } &\multicolumn{1}{c|}{(keV)} & \multicolumn{1}{c|}{($ \times 10^{-3}$)} \\


A  & 11-41    & 0.1-0.5 & 4.5\\
B  & 52-201   & 0.6-2.4 & 33.8\\ 
C  & 52-90    & 0.6-1.1 & 11.6\\
D  & 91-201   & 1.1-2.4 & 22.3\\
D1 & 91-150   & 1.1-1.8 & 20.3\\
D2 & 151-201  & 1.8-2.4 & 1.9\\

\hline
\end{tabular}
\label{tab3:3:1}
\end{table}


\subsubsection{Spectral fitting}
\label{sec:specfit}

The spectral data were analysed using {\scriptsize XSPEC} software. We were unable to produce formal fits to the spectrum with three free parameters (the temperature, $kT$, the column density, $n_{H}$, and the normalisation) using the $\chi^{2}$-minimisation  routines within {\scriptsize XSPEC} because of the poor signal to noise of the observation. 

In order to explore the parameter space and estimate limits on the temperature and column density, we chose representative values of $kT$ and $n_{H}$, based on previous {\it ROSAT\/} PSPC observations of the similar systems of near identical inclination in quiescence. The X-ray spectra of both HT Cas and Z Cha have been fitted with single component thermal Bremsstrahlung models, with temperatures of 2.4 keV in the case of HT Cas \cite{woodetal95a} and 4.4 keV for Z Cha \cite{vant97}. We also used the 6.0 keV temperature found for the optically thin plasma (e.g., Mewe et al. 1985) model fit to the EUV/X-ray spectrum of VW Hyi by Wheatley et al. (1996). Model fits to the X-ray spectra of various high-inclination non-magnetic CVs have yielded typical values of $n_{H} \sim 10^{20} - 10^{21}$ cm$^{-2}$ for the absorption, and we also adopted the value of $n_{H} = 10^{22}$ cm$^{-2}$ from the quiescent {\it HST\/} UV study of OY Car by Horne et al. (1994). Within {\scriptsize XSPEC}, we then allowed the normalisation to be a free parameter, freezing the chosen values of $kT$ and $n_{H}$. 

Table~4 gives values for the normalisation, emission measure and luminosities for the temperatures and absorbing columns under consideration. In general we found that lower temperatures produced higher values of $\chi^2$. 

We also attempted to fit the quiescent spectrum with a dual-absorption model, including both neutral and ionized absorbers. While the fits were formally as good as those discussed previously, the quality of the data does not allow us to constrain the contributions of each component.

A representative thermal Bremsstrahlung spectrum of $kT = 2.4$ keV and column densities of $n_{H} = 10^{20}$, $10^{21}$ and $10^{22}$ cm$^{-2}$, folded through the instrumental response, is shown in Figure~\ref{fig:quiespec}. The figure shows that a single-component column density of $10^{22}$ cm$^{-2}$ does not agree with the data for this temperature. Similar results are found for the other temperatures under consideration.

\begin{table*}
\begin{minipage}{120mm}
\center
\caption{{\small Table of quiescent luminosities and emission measures. The `normalisation' is the normalisation of the thermal Bremsstrahlung model in {\scriptsize XSPEC}, given by $K = (3.02 \times 10^{15})/4 \pi D^{2} EM$, where $D$ is the distance (in cm) and $EM$ is the emission measure. A bandpass of 0.001 - 20 keV is assumed to be bolometric.}}
\begin{tabular}{l l r l l r}
\hline

\multicolumn{1}{l}{kT} & \multicolumn{1}{l}{$n_{H}$} &\multicolumn{1}{l}{Normalisation} & \multicolumn{1}{l}{$EM \times 10^{53}$} & \multicolumn{1}{l}{Intrinsic luminosity} & \multicolumn{1}{l}{$\chi^2$}\\ 

\multicolumn{1}{l}{(keV)} & \multicolumn{1}{c}{($\times 10^{22}$ cm$^{-2}$)} & \multicolumn{1}{l}{($\times 10^{-4}$)} & \multicolumn{1}{l}{($\int n_{e}n_{I}dV$)} & \multicolumn{1}{l}{(erg s$^{-1}$)} & \multicolumn{1}{l}{($\nu = 10$)}\\ 

            &                 &                 &                        \\
 1.0 & 0.01 & $4.24 \pm 1.52$ & $1.13 \pm 0.41$ & $9.15 \times 10^{29}$ & 14.70 \\
 2.4 & 0.01 & $3.59 \pm 1.21$ & $0.96 \pm 0.32$ & $1.16 \times 10^{30}$ & 10.33 \\
 4.4 & 0.01 & $3.59 \pm 1.19$ & $0.96 \pm 0.32$ & $1.53 \times 10^{30}$ &  9.14 \\
 6.0 & 0.01 & $3.69 \pm 1.21$ & $0.98 \pm 0.32$ & $1.80 \times 10^{30}$ &  8.77 \\
     &      &                 &                 &                       &       \\
 1.0 & 0.1  & $7.31 \pm 2.41$ & $1.95 \pm 0.64$ & $1.58 \times 10^{30}$ & 11.22 \\
 2.4 & 0.1  & $5.32 \pm 1.73$ & $1.42 \pm 0.46$ & $1.72 \times 10^{30}$ &  9.03 \\
 4.4 & 0.1  & $5.09 \pm 1.64$ & $1.36 \pm 0.44$ & $2.17 \times 10^{30}$ &  8.51 \\
 6.0 & 0.1  & $5.16 \pm 1.66$ & $1.38 \pm 0.44$ & $2.52 \times 10^{30}$ &  8.30 \\
     &      &                 &                 &                       &       \\
 1.0 & 1.0  & $45.46 \pm 11.44$ & $12.11 \pm 3.05$ & $9.80 \times 10^{30}$ & 13.36 \\
 2.4 & 1.0  & $24.59 \pm 7.53$ & $6.55 \pm 2.01$ & $7.95 \times 10^{30}$ & 15.77 \\
 4.4 & 1.0  & $21.49 \pm 6.85$ & $5.73 \pm 1.83$ & $9.17 \times 10^{30}$ & 16.57 \\
 6.0 & 1.0  & $21.10 \pm 6.79$ & $5.62 \pm 1.81$ & $1.03 \times 10^{31}$ & 16.88 \\

\hline
\end{tabular}
\label{tab3:3:2:1}
\end{minipage}
\end{table*}

We used the values obtained for the normalisation as a result of the spectral investigation to calculate the corresponding emission measure. The emission measure is a robust physical determination because, as can be seen from Table ~\ref{tab3:3:2:1}, a factor of 100 change in the column density produces only a factor 7 change in the emission measure. The relatively low emission measures, of order $10^{53}$ cm$^{-3}$, in Table~\ref{tab3:3:2:1}, agree with the observations of other high-inclination systems in van Teeseling et al. \shortcite{vantetal96}.

Finally, within {\scriptsize XSPEC}, we estimated the luminosity of OY Car in quiescence (Table~\ref{tab3:3:2:1}), using the temperatures and column densities mentioned earlier. We used the same values for the distance and bandpass as for the superoutburst luminosity, Section~\ref{sec:sob_lum}. The quiescent X-ray luminosity cannot be constrained by this method to any better than $L_{Xq} \sim 10^{30 - 31}$ erg s$^{-1}$.


\section{The Superoutburst optical data}
\label{sec:sob_opt}


\subsection{Optical light curves}
\label{sec:optlcrvs}

\begin{figure}
\centerline{\psfig{figure=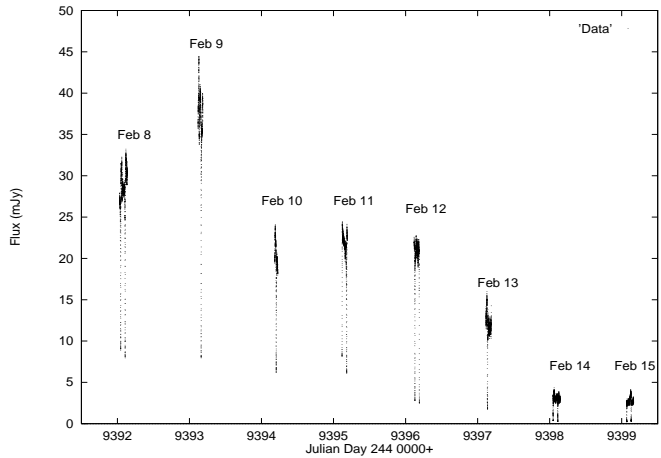,height=6cm,width=9cm,angle=0}}
\caption{All the optical light curves of OY Car obtained at the end of the superoutburst of Feb 1994. The decline in flux over the period in question is clearly seen. The observations taken on 1994 Feb 14 and 15 are virtually at the quiescent level.}\label{fig:all_lcrvs}
\end{figure}

\begin{figure*}
\begin{minipage}{14cm}
\centerline{\psfig{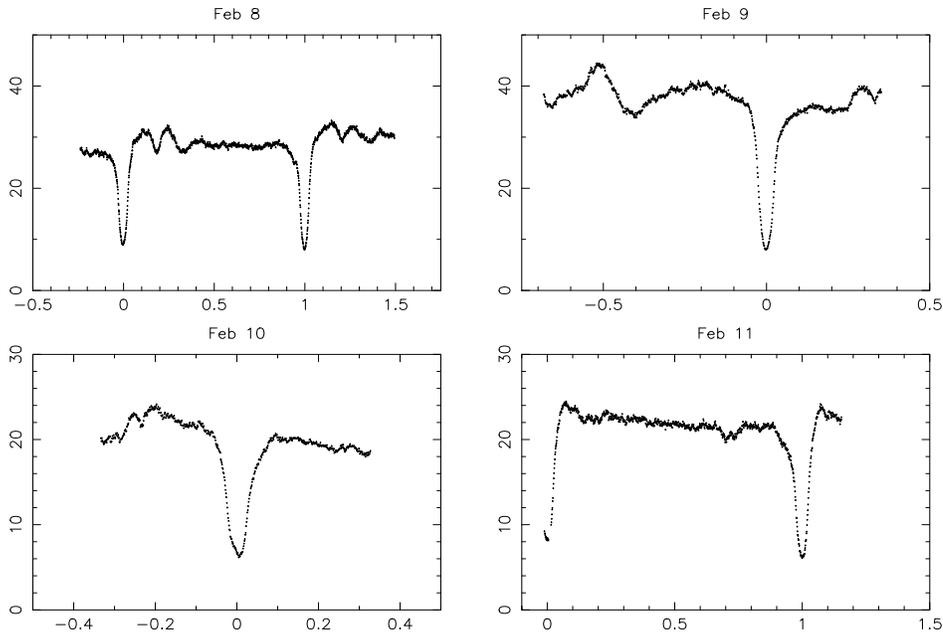}}
\caption{Four optical light curves of OY Car in superoutburst. The x-axis is orbital phase, the y-axis mJy. Note that the scale of the y-axis changes from Feb 9 to Feb 10.}\label{fig:first4}
\end{minipage}
\end{figure*}

\begin{figure*}
\begin{minipage}{14cm}
\centerline{\psfig{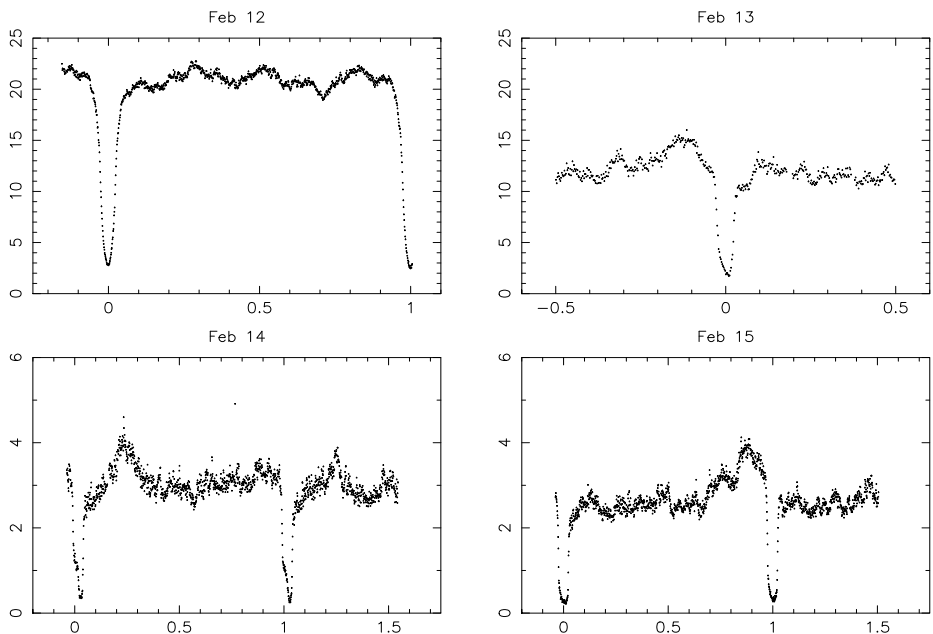}}
\caption{Four more optical light curves of OY Car on the decline from superoutburst. The x-and y-axes are as for figure~\ref{fig:first4}. Note that the scale of the y-axis has changed.}\label{fig:secnd4}
\end{minipage}
\end{figure*}

The overall behaviour of the system over the eight days of observations is shown in Figure~\ref{fig:all_lcrvs}. The decline in flux is marked over the period of the observations, with an especially sharp drop ($\sim 28 $ per cent) between the light curves of Feb 13 and Feb 14. Figures~\ref{fig:first4} and~\ref{fig:secnd4} show the individual optical light curves obtained over the period from 1994 Feb 8 to 1994 Feb 15, phased onto the orbital period of OY Car. 

The apparent rise in flux between the observations of 1994 Feb 8 and 1994 Feb 9 is real. Much of the excess flux in the light curve of Feb 9 appears to come from the large superhump feature. Orbital coverage is such that it is difficult to find a `continuum' level from which to make a quantitative estimate of the flux rise, but if we take the flux immediately after eclipse as a `continuum' level, then the rise in flux above the average flux level of the previous day is $\sim 20 $ per cent. 

The early eclipses appear to be roughly symmetric, but become increasing asymmetric as the decline progresses. On Feb 14, the superhump is at $\phi \sim 0.2$, and the eclipse shape is strikingly similar to that for quiescence; however, by Feb 15, the superhump is at $\phi \sim 0.8$, and the feature previously identified as the bright spot egress gives way again to a more round-bottomed eclipse. For the `most quiescent' light curve, that of Feb 14, we used the method of Wood, Irwin \& Pringle (1985) to measure the half-flux points of the eclipse, which in quiescence correspond to mid-ingress and egress of the white dwarf. We found a value of 255 s for the eclipse width, slightly longer than that of the white dwarf eclipse in quiescence (231s; Wood et al. \shortcite{woodetal89}), and a value that probably results from an over-estimation due to the superhump feature that is near the eclipse. 

The superhumps themselves show marked variations. On Feb 8 and Feb 9, the superhump is $\sim 0.25$ mag; it becomes less distinct on Feb 11 and Feb 12, and again becomes visible from Feb 13. Also visible are dips arising from self-absorption, similar to those seen in {\it HST} observations of OY Car in superoutburst by Billington et al. (1996). They can be attributed to obscuration of the central parts of the disc by the superhump material itself, explained by Billington et al. (1996) as time dependent changes in the thickness of the disc at the outer edge. This azimuthal asymmetry is distinct from the wholescale physical flaring of the disc. The best example of the dip phenomenon is in the light curve of Feb 8, where it can be seen to occur in the superhump over two consecutive orbital periods. The dips also come and go over the course of our observations, being visible in the light curves of Feb 8, 9, 13, and 15, but not visible in the light curves of Feb 10, 11, 12 and 14. The superhump peaks are difficult to define when the dips are significant. Calculations using an approximate superhump ephemeris reveal that the superhumps visible in the light curves of Feb 14 and Feb 15 are late superhumps, shifted in phase from the normal superhumps by 180$^{\circ}$. Both the late superhump ($\phi \sim 0.25$) and a faint quiescent orbital hump ($\phi \sim 0.85$) can be seen in the light curve of Feb 14. The late superhump visible in the Feb 15 light curve is enhanced in intensity due to a coincidence with the quiescent orbital hump.


\subsection{The superoutburst disc}


\subsubsection{Introduction}

\label{sec:disc_intro}

To investigate the surface brightness distribution of the disc during the superoutburst, we used the eclipse mapping method developed by Horne (1985). The original method rests on three basic assumptions:

\begin{itemize}

\item The surface of the secondary star is given by its Roche potential,
\item The accretion disc lies flat in the orbital plane, and,
\item The emitted radiation is independent of the orbital phase, i.e., the observed surface brightness distribution of the disc remains fixed with respect to the binary.

\end{itemize}

The method thus allows deconvolution of the eclipse profile. This is achieved by comparing, using maximum entropy methods, the observed intensity distribution with a default distribution specified by the above items and the parameters of the system. The brightness of the observed disc is then determined, subject to two constraints:

\begin{itemize}

\item Agreement between the calculated and observed light curves is achieved by the use of the $\chi^{2}$ statistic, and,

\item To allow a choice of one of the many possible disc light distributions, the disc template map and light curve must be as close to axisymmetric as possible.

\end{itemize}

The eclipse mapping program used in our analysis is an extension of Horne's original code that allows a flared disc, i.e., one that increases in thickness outward from the centre, but is not concave. For this Section and the remainder of the paper, we will use $\alpha$ to denote the opening angle of the disc. Thus $\alpha/2$ is the flare half angle. That the line of sight must intercept the disc surface constrains the allowed flare angles to $\alpha/2 < 90 - i$, which in our case is $< 7^{\circ}$.

\subsubsection{Method}

\label{sec:discmethod}

The disc dominates the flux at the optical bandpass used, thus the contribution from the secondary will be negligible. In choosing which light curves to map, we avoided those in which eclipses were contaminated with superhump light, as the presence of this excess light can affect the disc light distribution and cause spurious artifacts, especially if near phase zero. For our data set, it is clear that the light curves of 1994 Feb 8, 10, 11, 13, and 15 exhibit signs of superhump light contamination of the eclipse phase. These light curves also show dips, discussed above; these would render our eclipse mapping program invalid, as in this case $\alpha/2 > 90 - i$.  We thus chose to map the light curves of 1994 Feb 9 and 12. We also attempted to map the light curve of Feb 14, but the poor signal to noise of the observation made this difficult and introduced too much uncertainty for a realistic disc map.

The February 9 and February 12 data were each divided into 40 bins for orbital phase $-0.2 \leq \phi \leq 0.2$. We found that simple Poissonian statistics produced uncertainties that were small when compared to the intrinsic flickering of the data, so we adopted a constant percentage error derived from the ratio of the flickering to the average flux out of eclipse, thus giving what we regard as a more realistic measure of the errors. This constant percentage error was different for each data set, as the mean out of eclipse flux has dropped by a factor $\sim 40$ percent in the 3 day period between the light curves. We used a mass ratio $q = 0.102$, a distance $D = 82$ pc, and inclination $i = 83.3^{\circ}$ as standard parameters, all from Wood et al. (1989), and a Cartesian grid of $69 \times 69$ pixels, with the white dwarf at the centre, to cover the accretion disc plane.

We initially started mapping the light curves with a flat disc. This produced maps with a pronounced front-back asymmetry and a high value of $\chi^{2}$, as would be expected if the flare had been ignored (Wood 1994). We then began mapping the data with a flared disc, gradually incrementing the value of the disc opening angle, $\alpha$, from the flat case. This produced an immediate improvement in the front-back asymmetry and reduced the value of $\chi^{2}$. By using a combination of the front-back asymmetry and $\chi^{2}$ diagnostics, we arrived at acceptable fits to each light curve. The corresponding disc opening angles were $\alpha = 11^{\circ}$ for the 1994 Feb 9 light curve, and $\alpha = 10^{\circ}$ for the Feb 12 light curve, respectively.  The resulting contour maps and associated radial brightness temperature distribution in the discs of February 9 and February 12 can be seen in Figures~\ref{fig:feb09cont}, ~\ref{fig:feb09britemp}, ~\ref{fig:feb12cont}, and~\ref{fig:feb12britemp}, respectively. For comparison, the disc map and temperature distribution obtained when the light curve of Feb 12 is fitted with a flat disc are shown in Figures~\ref{fig:fdcont} and~\ref{fig:fdbritemp}. Figure~\ref{fig:datamdls} shows the light curves calculated from the flat and flared disc maps, plotted with the actual data. It is obvious that the flared disc is a far better fit to the data.

\begin{figure}
\centerline{\psfig{figure=fig8.eps,height=6.0cm,width=8.5cm,angle=270}}
\caption{The reconstructed surface brightness in the disc of February 9. The flare angle is $\theta = 11^{\circ}$. The contour scale is logarithmic, with a contour interval of 0.2 decades.}\label{fig:feb09cont}
\end{figure}

\begin{figure}
\centerline{\psfig{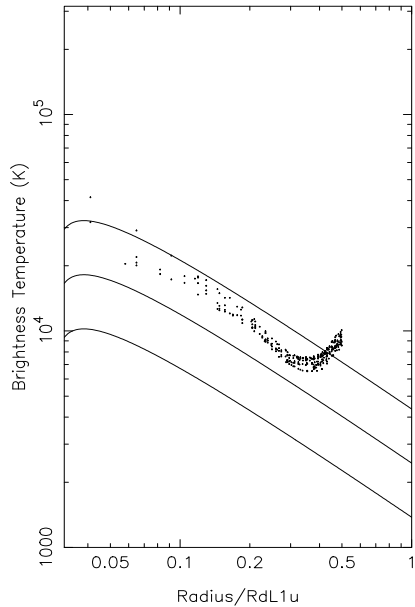}}
\caption{The brightness temperature distribution in the Feb 9 disc. Models of steady state optically thick for, from top to bottom: $\dot{M} = 10^{-9}$, $10^{-10}$, and $10^{-11}$ M$_{\odot}$ yr$^{-1}$, are also shown.}\label{fig:feb09britemp}
\end{figure}


\section{Discussion}
\label{sec:res_disc}


\subsection{OY Car and VW Hyi in the EUV and X-ray}

In this section, we discuss X-ray and EUV observations of VW Hyi and OY Car in order to address the question of whether or not the behaviour of OY Car in this wavelength range is typical for an SU UMa-type dwarf nova. Interpretation of the X-ray data from these two systems involves the consideration of several important interconnected factors:

\begin{itemize}
\item Firstly, OY Car is an eclipsing system, whereas VW Hyi is not. 

\item Secondly, the absorption to the X-ray source must be taken into account. The absorption can have a maximum of three possible components. First and foremost there is the {\it interstellar absorption}, which for VW Hyi is the lowest of all dwarf novae, at $n_{H} \sim 6 \times 10^{17}$ cm$^{-2}$ \cite{bellonietal91}, but which is unknown in the case of OY Car. Secondly, there is the {\it local absorption\/} due to the presence of the `iron curtain' of material which veils the white dwarf in quiescence \cite{hornetal94} and possibly outburst. Lastly, in the case of high inclination systems in outburst and superoutburst, there is the effect of {\it disc edge absorption}, a consequence of the material of the disc edge obscuring the central regions while flared. This last effect can be seen in the dips of the optical light curves in this paper, while Billington et al. (1996) find dips in {\it HST} lightcurves of OY Car in superoutburst that are coincident with the optical superhump maximum. It is not known whether the `iron curtain' phenomenon is connected to the disc, although the fact that to date, the `iron curtain' has only been seen in high inclination systems certainly suggests that this is so. There is no known way of disentangling the effect of absorption due to the material of an `iron curtain' from that of a disc seen nearly edge-on.

\item Lastly, the spectral responses of the various X-ray satellites are all different -- the responses depend on the instrument and detector used aboard the observing satellite. Table~\ref{tab5:1:1} summarises the instruments (and their respective spectral ranges) aboard the EUV/X-ray satellites relevant to this discussion, viz, {\it EUVE, EXOSAT, ROSAT}, and {\it GINGA}. A comparative illustration of the effective areas and spectral responses of the instruments aboard {\it EXOSAT} and {\it ROSAT} can be found in Figure 9 of Wheatley et al. (1996).
\end{itemize}

\begin{figure}
\centerline{\psfig{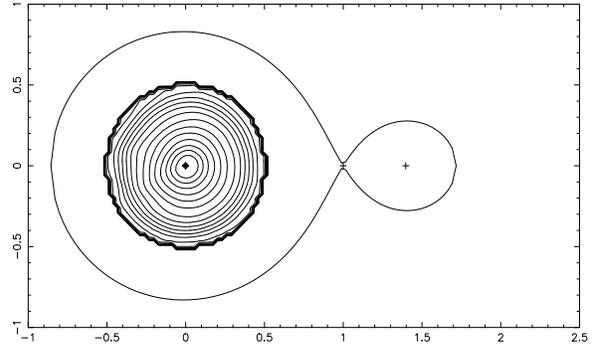}}
\caption{The reconstructed surface brightness in the disc of February 12. The flare angle is $\theta = 10^{\circ}$. The contour scale is logarithmic, with a contour interval of 0.2 decades.}\label{fig:feb12cont}
\end{figure}

\begin{figure}
\centerline{\psfig{figure=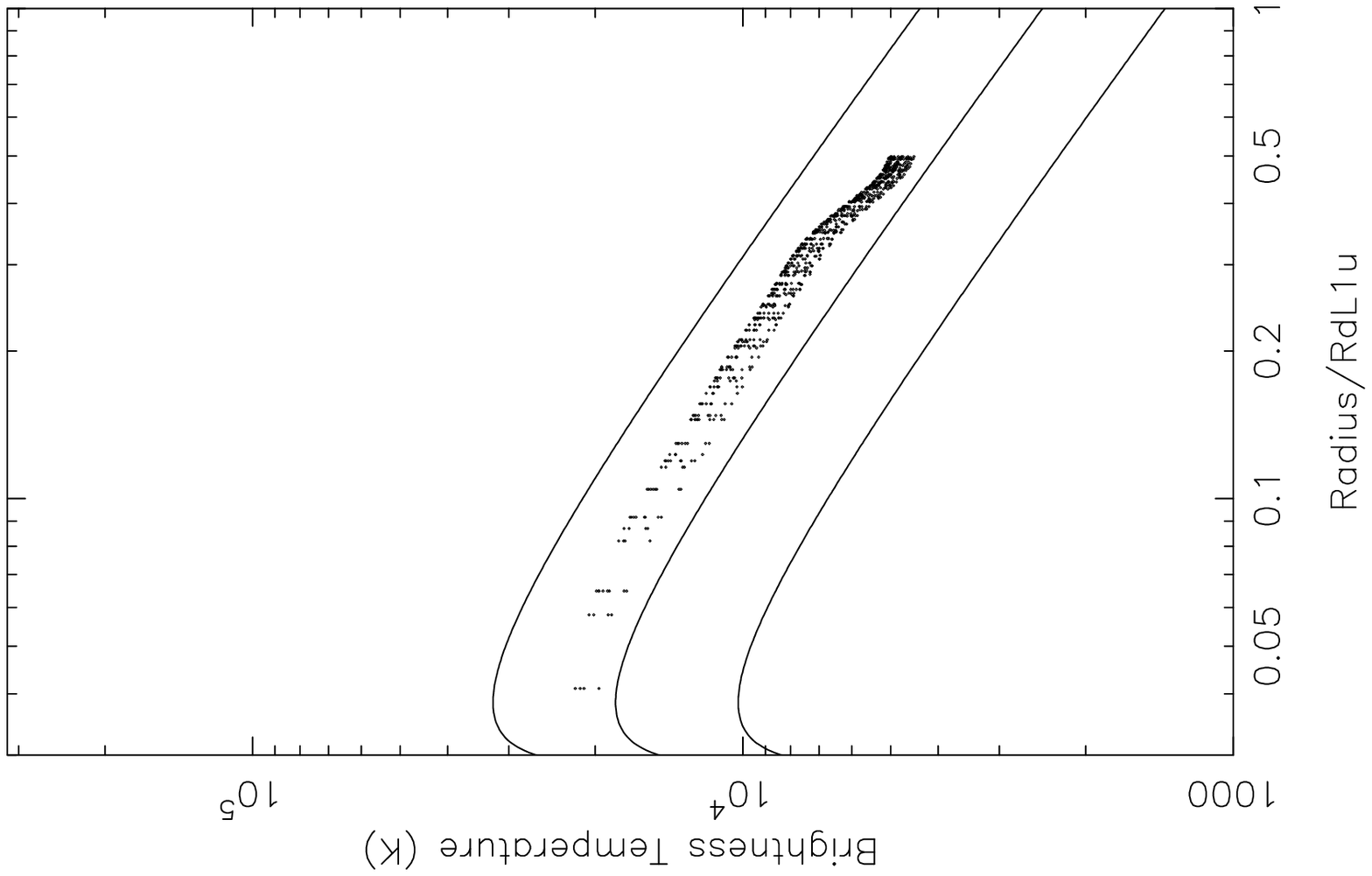,height=8.0cm,width=5cm,angle=270}}
\caption{The brightness temperature distribution in the Feb 12 disc. Models of steady state optically thick for, from top to bottom: $\dot{M} = 10^{-9}$, $10^{-10}$, and $10^{-11}$ M$_{\odot}$ yr$^{-1}$, are also shown.}\label{fig:feb12britemp}
\end{figure}

\begin{figure}
\centerline{\psfig{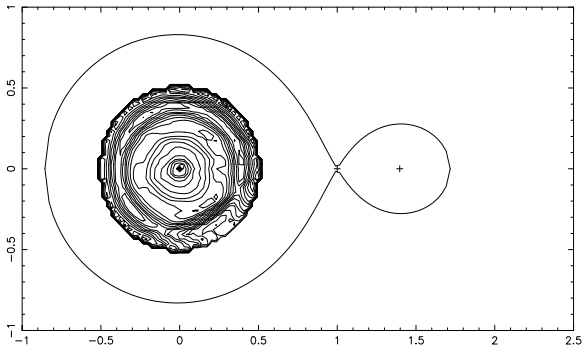}}
\caption{For comparison, this is the reconstructed surface brightness in the disc of February 12 using a flat disc model. The contour scale is logarithmic, with a contour interval of 0.2 decades.}
\label{fig:fdcont}
\end{figure}

\begin{figure}
\centerline{\psfig{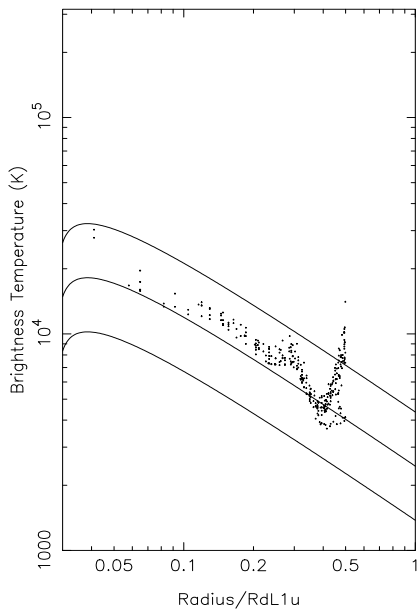}}
\caption{The corresponding brightness temperature distribution in the Feb 12 disc when fitted with a flat disc model. Models of steady state optically thick for, from top to bottom: $\dot{M} = 10^{-9}$, $10^{-10}$, and $10^{-11}$ M$_{\odot}$ yr$^{-1}$, are also shown.}\label{fig:fdbritemp}
\end{figure}

VW Hyi is perhaps the best studied low inclination SU UMa dwarf nova, bright in the optical ($m_{v}$(max) $\sim 9.5$, $m_{v}$(min) $\sim 13.3$), and observed to undergo regular outbursts ($T_{n} \sim 30$d; Ritter 1997), and superoutbursts ($T_{s} \sim 180$d; Ritter 1997). Its brightness is to some extent due to the low interstellar absorption to the system; it has thus been well studied in all wavelengths, including the X-ray. 

\begin{table}
\center
\caption{{\small Instruments, filters, and spectral energy ranges of recent EUV and X-ray satellites.}}
\begin{tabular}{|l|l|l|}
\hline

\multicolumn{1}{|c}{Satellite} & \multicolumn{1}{|c|}{Instrument name} &\multicolumn{1}{c|}{Range (keV)}\\ 

             &                    &               \\
{\it EUVE}   & Lex/B scanner      & 0.07 - 0.25   \\
             & Al/Ti/C scanner    & 0.05 - 0.08   \\
             &                    &               \\
             & SW spectrometer    & 0.07 - 0.18   \\
             & MW spectrometer    & 0.03 - 0.09   \\
             & LW spectrometer    & 0.02 - 0.04   \\
             &                    &               \\
{\it EXOSAT} & Low Energy (LE)    &               \\
             & 3000-Lexan         & 0.05 - 1.77   \\
             & 4000-Lexan         & 0.07 - 1.77   \\ 
             & Al-Par             & 0.04 - 1.77   \\
             & Medium energy (ME) & 1 - 20        \\
             &                    & 5 - 50        \\
             &                    &               \\
{\it ROSAT}  & WFC                &               \\
             & S1a                & 0.09 - 0.21   \\
             & S2a                & 0.06 - 0.11   \\
             & PSPC/HRI           & 0.1 - 2.5     \\
             &                    &               \\
{\it GINGA}  & Large Area Counter & 2 - 30        \\
               
\hline
\end{tabular}
\label{tab5:1:1}
\end{table}


\begin{figure*}
\centerline{\psfig{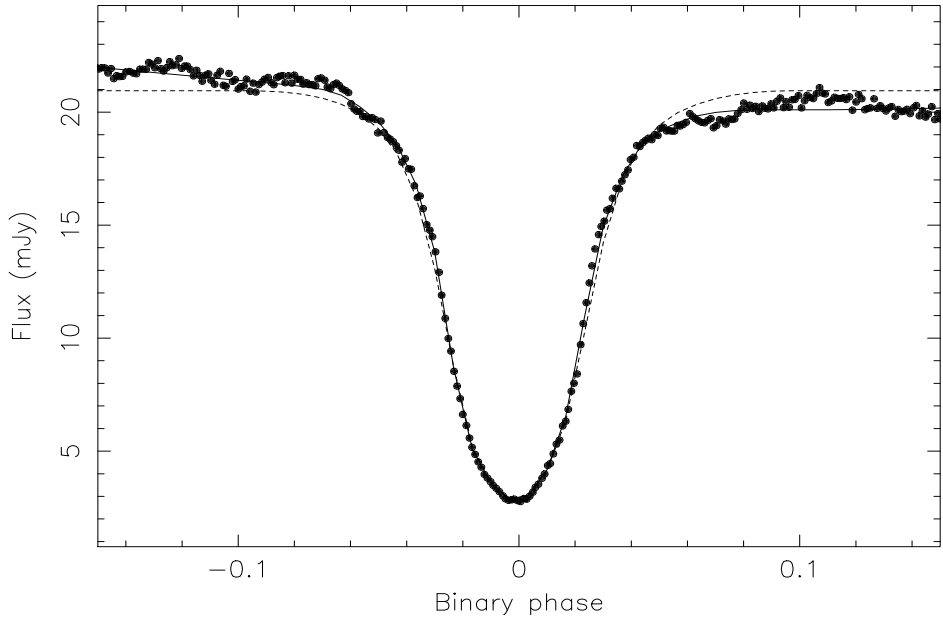}}
\caption{A comparison of the light curves calculated from the reconstructed disc map of Feb 12 when modelled with a flat disc (dashed) and a flared disc of opening angle $\alpha/2 = 5^{\circ}$ (solid).}\label{fig:datamdls}
\end{figure*}

\subsubsection{Summary of observations to date}

\paragraph{EUV data}

VW Hyi is the only dwarf nova to be detected in quiescence by the {\it ROSAT\/} WFC, likely a consequence of the low interstellar column in this direction. VW Hyi was not observed above background in quiescence by either of the detectors aboard {\it EUVE} during the all-sky survey phase. 

VW Hyi has also been detected in outburst with the {\it ROSAT\/} WFC, although this was during the all-sky survey phase. VW Hyi has been detected in superoutburst with the {\it EUVE}. Mauche (1996) shows different count rates depending on the spectrometer, with the count rate rising for the duration of the observation across all three detectors. The corresponding EUV spectrum of VW Hyi is shown to peak longward of 0.07 keV, and does not extend shortward of 0.12 keV. In Section~\ref{sec:obs_dat} we show that OY Car was not detected by the {\it ROSAT\/} WFC during the 1994 Jan superoutburst. 

\paragraph{X-ray data}

Fits to the quiescent X-ray spectrum of VW Hyi have produced various values for the temperature of the emitting gas, typically yielding $kT < 10$ keV, depending on the spectral response of the satellite involved. Belloni et al. (1991), produced a quiescent temperature of $kT = 2.17 \pm 0.15$ keV when the {\it ROSAT\/} PSPC observation was fitted with a single-component Raymond-Smith optically thin thermal plasma with line emission at $\sim 1$ keV. Wheatley et al. (1996) extended the analysis further by combining the {\it ROSAT\/} PSPC observation of Belloni et al. with {\it ROSAT\/} WFC EUV data at the low end and {\it GINGA} hard X-ray data at the high end, giving unprecedented wavelength coverage of the high-energy spectrum of VW Hyi. Wheatley et al. fitted this combined spectrum with single and two-temperature Mewe optically thin plasma models, with and without line emission. They found that the dominant continuum temperature is $\sim 6$ keV for both model types, although they were unable to use the data to distinguish between model types. For the two-temperature case, the components are at temperatures of $\sim 6.0$ keV and $\sim 0.7$ keV.


Our {\it ROSAT} observation of OY Car in quiescence does not contain enough information to do more than constrain the X-ray spectrum and the absorption to the X-ray source. We can however conclude that the quiescent spectrum of OY Car is produced by a boundary layer at a typical temperature for this type of object: i.e., $kT < 10$ keV.

\begin{table}
\begin{minipage}{85mm}
\center
\caption{Comparison of the EUV observations of OY Car and VW Hyi.}
\begin{tabular}{l|l|l}
\hline

\multicolumn{1}{|l}{} & \multicolumn{1}{|l}{OY Car} &\multicolumn{1}{|l|}{VW Hyi} \\

{\it ROSAT} WFC  &               &                                  \\       
                 &               &                                  \\
Quiescence       & No detection\footnote{$1.5 \times 10^{-3}$ cts s$^{-1}$ upper limit}  & $18.7 \pm 2.7$ cts s$^{-1}$ \\
Outburst         & Unobserved    & $ 0.08 \pm 0.04$ cts s$^{-1}$    \\
Superoutburst    & No detection\footnote{$1.2 \times 10^{-3}$ cts s$^{-1}$ upper limit}  & Unobserved                       \\
		 &               &                                  \\

{\it EUVE}       &               &                                  \\
                 &               &                                  \\
Quiescence       & Unobserved    & No detection                     \\
Outburst         & Unobserved    & $\sim 0.09$ cts s$^{-1}$\footnote{In the AL/Ti/C detector during the all-sky survey phase}         \\ 
Superoutburst    & Unobserved    & $\sim 1.0$ cts s$^{-1}$ (SW)     \\
                 &               & $\sim 4.0$ cts s$^{-1}$ (MW)     \\
                 &               & $\sim 2.0$ cts s$^{-1}$ (LW)     \\
                 &               &                                  \\
\hline
\end{tabular}
\label{tab6:1}
\end{minipage}
\end{table}

OY Car and VW Hyi have been observed in an outburst state by both {\it EXOSAT} and {\it ROSAT}, although in the case of OY Car a normal outburst has never been caught. There is an increase in the soft X-ray flux from both systems during superoutburst (e.g., van der Woerd, Heise \& Bateson, 1986; Naylor et al. 1988), but the actual scale of the increase differs --- the increase is a factor of $\sim 100$ for VW Hyi, whereas for OY Car it rises only by a factor $\sim 3$. 

According to van Teeseling, Verbunt \& Heise (1993), the {\it EXOSAT}-observed superoutburst X-ray spectrum of VW Hyi can be modelled in terms of either a single component optically thin spectrum with 0.043 keV $\leq kT \leq$ 0.43 keV, or a combination of a hot, optically thin component at $kT >$ 0.43 keV (which they tentatively identify with the 2.17 keV quiescent spectrum found by Belloni et al. 1991) and a cooler, optically thick component at 0.034 keV $< kT <$ 0.042 keV. Wheatley et al. (1996) prefer the optically thin plus blackbody component interpretation for the spectrum of VW Hyi in both outburst and superoutburst. They suggest the superoutburst spectrum is a result of a drop of a factor $\sim 3-4$ in the hot quiescent optically thin component and the rise of a very soft blackbody component that is easily detected by the {\it EXOSAT} filters but not by the {\it ROSAT} instruments. In other words, the quiescent and outburst spectra can both be explained in the context of a variation in the relative strengths of just the two components. They find no evidence for spectral change (from thermal Bremsstrahlung to blackbody, for example) during outburst observations. 

OY Car has now been observed by both {\it EXOSAT} and {\it ROSAT} in superoutburst. The {\it EXOSAT} observation by Naylor et al. (1988) suggested that the X-rays in the 1985 May superoutburst came from a large optically thin corona at $kT \sim 0.43$ keV, producing a luminosity of $L_{{X}_{SOB}} \sim 10^{31} - 10^{32}$ erg s$^{-1}$. Our {\it ROSAT} superoutburst observation of OY Car has a count rate comparable to the quiescent count rate, taking into account the fact that different detectors were used. This suggests that the {\it ROSAT} instruments are detecting a greater part of the hard, optically thin component than the corresponding detectors aboard {\it EXOSAT}.

Table~6 summarises the EUV and X-ray observations of OY Car and VW Hyi to date.


\subsubsection{So, how similar are OY Car and VW Hyi?}

The approximate X-ray luminosity of each system in each state is shown in Table~\ref{tab6:2}, from which it can be seen that the luminosity of OY Car is consistently the lower of the two. But this could be explained qualitatively in terms of the three differentiating factors, namely: the satellite spectral responses, the inclination and the absorption.
 
Satellite spectral responses can and do confuse the issue. Wheatley et al. (1996) describe the evolution of the X-ray spectrum of VW Hyi from quiescence to outburst as the appearance of a very soft blackbody component accompanied by a change in the relative strengths of two optically thin components. In quiescence, the optically thin components dominate, while in outburst, the very soft blackbody component rises to contribute most of the flux outside the {\it ROSAT\/} bandpass. This component has disappeared by the end of the outburst. Assuming that the same components occur in OY Car, we must bear in mind that only the optically thin components can contribute to the {\it ROSAT} bandpass. This suggests that we are underestimating the total high-energy output of OY Car because neither the {\it ROSAT\/} HRI or the PSPC are sensitive to the lower energy ranges.

Additionally, the inclination of the system is crucial. The inclination of VW Hyi simply allows more of the X-ray and EUV emitting boundary layer to be seen directly. For OY Car however, although the boundary layer can be seen in quiescence, during outburst/superoutburst there is no direct line of sight because the disc has flared up and the inclination is such that it blocks our view. There is increased absorption seen in OY Car as there is far more material in the orbital plane than there is in VW Hyi. The lack of X-ray eclipse, together with the evidence of a disc with an opening angle of $10^{\circ}$ three days after the end of the superoutburst, illustrate this effect. The significantly lower X-ray luminosity in OY Car during superoutburst is another consequence of this increased absorption.



\begin{table*}
\begin{minipage}{130mm}
\center
\caption{Intrinsic luminosity values for OY Car and VW Hyi. Very approximate, uncorrected for inclination.}
\begin{tabular}{l|l|l|l|l|l|l}
\hline

\multicolumn{1}{|l}{System} & \multicolumn{1}{|l}{State} &\multicolumn{1}{|l|}{$L_{X}$} & \multicolumn{1}{|l}{$L_{UV}$\footnote{Calculated using $F_{UV} = 718f_{1460} + 340f_{1800} + 540f_{2140} + 490f_{2880}$}} & \multicolumn{1}{|l}{$L_{opt+UV}$\footnote{Calculated using $F_{opt+UV}  =  718f_{1460} + 340f_{1800} + 540f_{2140} + 1680f_{2880} + 2620f_{5500}$. Method used by van Teeseling et al. (1996), consisting of summing the average fluxes at $1460\AA,\ 1800\AA,\ 2140\AA,\ 2880\AA$, and $5500\AA$, in erg s$^{-1} \AA^{-1}$, multiplied by their weights }} & \multicolumn{1}{|l}{$L_{X}/L_{UV}$} & \multicolumn{1}{|l}{$L_{X}/L_{opt+UV}$} \\

\multicolumn{1}{|l}{} & \multicolumn{1}{|l}{} &\multicolumn{1}{|l|}{(erg s$^{-1}$)} & \multicolumn{1}{|l}{(erg s$^{-1}$)} & \multicolumn{1}{|l}{(erg s$^{-1}$)} & \multicolumn{1}{|l}{} & \multicolumn{1}{|l}{} \\

      &    &     &     &    &   &   \\

OY Car & quiescent     & $ \sim 10^{30}$ & $1.7 \times 10^{31}$ & $2.5 \times 10^{31}$ & 0.06 & 0.04 \\
       & superoutburst & $\sim 2 \times 10^{32}$ & $1.4 \times 10^{32}$ & $3.5 \times 10^{32}$ & $0.07 -1.43$ & $0.03 - 0.57$ \\

      &    &     &     &    &   &    \\

VW Hyi & quiescent     & $ \sim 10^{31}$ & $9.1 \times 10^{31}$ & $1.4 \times 10^{32}$ & 0.11 & 0.07 \\
       & superoutburst & $ \sim 10^{34}$ & $1.2 \times 10^{34}$ & $1.6 \times 10^{34}$ & 0.83 & 0.63 \\

\hline
\end{tabular}
\label{tab6:2}
\end{minipage}
\end{table*}


\subsection{Superoutburst: flared discs and the lack of X-ray eclipse}

\subsubsection{Flare angles and temperature distributions}


In the context of accretion disc mapping, the motivation for considering flared discs has arisen from the poor fitting, and the occurrence of spurious structure, that have resulted from flat disc model fits to outburst and superoutburst light curves. During outbursts of Z Cha (Warner \& O'Donoghue, 1988; Robinson et al. 1995) and OY Car (Rutten et al. 1992a), bright rings around the outer edge of the disc and/or front-back asymmetries were present in the reconstructed disc when the light curve was mapped with a flat disc. There are two interpretations for this phenomenon: either the observed features are due to an uneclipsed light source (e.g., the secondary star, or a large disc), or the disc is exhibiting a physical flare. See the review by Wood \shortcite{wood94} for further discussion. 

Mapping light curves with flared discs has been attempted at normal outburst maximum for both OY Car and Z Cha. Rutten et al. (1992a) briefly considered a flared disc in their optical examination of OY Car at normal outburst peak. They used small opening angles and did obtain a satisfactory fit to the light curve, but this caused a front-back asymmetry to develop. Rutten et al. (1992a) prefer the uneclipsed light source interpretation, finally eliminating the spurious features by subtracting 15 per cent of the flux before mapping. 

However, the {\it HST} light curve of Z Cha at normal outburst maximum was mapped with a disc of opening angle $\alpha = 16^{\circ}$ by Robinson et al. \shortcite{robinsonetal95}. The fact that by mid-decline, the light curve could be modelled by an axisymmetric flat disc, lends weight to the interpretation of these data at outburst maximum as due to a flared disc.

We demonstrated in Section~\ref{sec:discmethod} that a disc with a substantial flare exists in OY Car three days into the decline from superoutburst. These are the first maps of a disc in the superoutburst state that clearly show such flaring. 

By way of illustrating the differences between observations and theory, we can calculate the theoretical flare angles and compare these to the results from eclipse mapping.

The {\it steady state} theoretical flare angle is given by (Shakura \& Sunyaev 1973; Frank, King \& Raine, 1985)

\begin{eqnarray*}
\frac{h_{d}}{r_{d}}  & = & 1.72 \times 10^{-2} \alpha_{v}^{-\frac{1}{10}} M_{WD}^{\frac{3}{8}} (r_{d} \times 69.6)^{\frac{1}{8}} \ldots \\
 & & \ldots \left(\frac{\dot{M}}{1.587 \times 10^{-10}}\right)^{\frac{3}{20}} \left\{ 1- \left[ \frac{R_{WD}}{r_{d}} \right]^{\frac{1}{2}} \right\}^{\frac{3}{5}}
\end{eqnarray*}

\noindent Here the equation is normalised to solar values, and $\alpha_{v}$ is the viscosity. Assuming that the disc radius is approximately equal to the tidal radius ($r_{d} \sim r_{t}$), $r_{t}$ is approximated by 

\[
r_{t} = r_{d} = \frac{0.6 a}{1+q}
\]

\noindent (Warner 1995; after Paczynsky 1977), where $a$ is the binary separation and $q$ is the mass ratio. 

Using $M_{WD} = 0.685 M_{\odot}$, $a = 0.608 R_{\odot}$, $q = 0.102$ (Wood et al. 1989), and assuming $\alpha_{v} \sim 1.0$, we find a {\it steady state} disc flare angle of $\alpha \simeq 2.6^{\circ}$.

Alternatively, if we assume that the disc is in a {\it non-steady state}, and that the disc includes radiative and convective zones, then 

\[
\frac{h_{d}}{r_{d}} \sim 0.038 \left( {\frac{\dot{M}}{1.587 \times 10^{-10}}} \right)^{\frac{3}{20}} 
\]

\noindent (adapted from Smak 1992). Adopting $\dot{M} \sim 4 \times 10^{-10} M_{\odot}$ yr$^{-1}$ (from Figure~\ref{fig:feb12britemp}), gives $\alpha \simeq  5^{\circ}$. 

Thus the flare opening angle of OY Car on superoutburst decline 
is considerably larger than that predicted by either steady state or non-steady state theory.

We find a flare angle of $\alpha \simeq 10^{\circ}$ for the disc of Feb 12; however we note that limb darkening has not been taken into account. The first steps towards including limb darkening in eclipse mapping have recently been taken by Robinson and collaborators, using the {\it HST} light curve previously mapped with a flared disc of opening angle $\alpha/2 = 8^{\circ}$ (Robinson, Wood and Wade 1999). The net result of including the limb darkening in their analysis of this UV data set is to reduce the opening angle of the disc to $\alpha/2 = 6^{\circ}$, a value that is still between $2^{\circ}$ and $3^{\circ}$ larger than that given by disc model atmospheres. We note that limb darkening will have less of an effect on our $B$ band optical data. 

Figure~\ref{fig:feb09britemp} shows that the slope of the disc brightness temperature distribution on Feb 9 appears only approximately compatible with the theoretical steady state solutions. The overall rate of mass flow rate is slightly larger than that for the disc of Feb 12. 

The turn-up at the edge of the brightness temperature distribution on Feb 9 is similar to that seen in disc maps of Z Cha ( Warner \& O'Donoghue 1988). Physical explanation of this effect requires the presence of a ring at the outer edge. We would hesitate to invoke an explanation that requires a ring if the light curves were acquired in quiescence, but numerical simulations of superoutburst discs by Whitehurst (1988a, 1988b) and Whitehurst \& King (1991) suggest that rings exist at the outer edges of the disc in superoutburst, and indeed, are the most likely explanation for the superhump phenomenon. Unfortunately, in this case, we cannot say definitively whether the ring is real or an artifact from the modelling process.

In contrast, the observed slope of the disc brightness temperature distribution on Feb 12 (Figure~\ref{fig:feb12britemp}) is quite compatible with the overplotted theoretical curves of optically thick, steady state discs. The average rate of mass flow through the disc at this stage in the decline appears to be $\sim 4 \times 10^{-10}$ M$_{\odot}$ yr$^{-1}$. There is no evidence for a ring at the outer edge of the disc at this stage in the superoutburst decline.

This can be compared with a normal outburst peak mass flow rate of $\sim 1 \times 10^{-9} M_{\odot}$ yr$^{-1}$ for OY Car, again with a relatively good fit to steady state theoretical disc curves (Rutten et al. 1992a). We note the fact that our disc mass flow rate is lower than that found at normal outburst maximum. However, the data presented here were obtained well into the decline from superoutburst --- the flux from the system is virtually at the quiescent level two days after the acquisition of these data.


\subsubsection{Tying it all together}

A self-consistent picture of OY Car in superoutburst emerges with the X-ray and optical data presented in this paper. The lack of X-ray eclipse affirms the conclusions of the {\it EXOSAT} observation of the 1985 May superoutburst by Naylor et al (1988), suggesting that their extended, `coronal' X-ray source may well be the rule and not the exception when superoutbursts occur in high inclination systems. Our value for the X-ray luminosity in superoutburst is also in agreement with the value of $L_{X} \sim 10^{31} - 10^{32}$ erg s$^{-1}$, found by Naylor et al. (1988). This is up to 100 times less luminous than VW Hyi in superoutburst.

Naylor et al. (1987, 1988) used contemporaneous observations in other wavelengths to suggest that the disc of OY Car had considerable vertical structure during the 1985 May superoutburst, and that the increased local absorption that resulted had the effect of shielding the boundary layer from view at all orbital phases. This also results in a decrease in the detected X-ray flux, leading to false conclusions being drawn regarding the X-ray luminosity of the system and a very low $L_{X}/L_{opt+UV}$ ratio. The optical data presented here show for the first time how flared the disc is even three days into the superoutburst decline. It is not difficult to envision even larger opening angles during superoutburst maximum, but it is unclear whether the disc is thickened at the edge, or nearer the centre; even more perturbing is the lack of a simple mechanism to support this structure.


\subsection{Quiescence: Constraining the absorption}

In terms of the emission measure and hardness ratio, OY Car is similar to the other high inclination dwarf novae investigated in van Teeseling \& Verbunt (1994) and van Teeseling et al. (1996): the volume of the X-ray emitting source is small compared to that in lower inclination systems.

The quiescent X-ray luminosity of OY Car appears comparable to both Z Cha (bolometric $L_{{X}_{q}} \sim 2.5 \times 10^{30}$ erg s$^{-1}$; van Teeseling, 1997) and HT Cas in its unusually low state (bolometric $L_{{X}_{q}} \sim 5 \times 10^{30}$ erg s$^{-1}$; Wood et al. 1995a), but is an order of magnitude below the X-ray luminosity from the {\it ASCA} observation of HT Cas in the normal quiescent state ($L_{{X}_{q}} \sim 2.2 \times 10^{31}$ erg s$^{-1}$; Mukai et al. 1997). Cordova \& Mason (1984a) quote typical CV X-ray luminosities of $\sim 10^{31}$ erg s$^{-1}$ in quiescence, although it must be remembered that the proportion of eclipsing high inclination CVs is small. 

If the X-rays do come from the boundary layer, then assuming the steady state case, (i.e., $L_{bl} = \frac{1}{2} GM_{WD} \dot{M}/R_{WD}$); for the case of OY Car in quiescence, $\dot{M}$ is $3 \times 10^{-13} M_{\sun}$ yr$^{-1}$, an exceptionally low value. Wood et al (1995a) find a mass transfer rate onto the white dwarf of $\dot{M} \sim 1 \times 10^{-12} M_{\sun}$ yr$^{-1}$ for the low state of HT Cas. 

One possible explanation for such a low X-ray luminosity is that the white dwarf in these systems is spinning rapidly. This is not inconceivable, given that Sion et al. (1995a), using high resolution {\it HST} spectra, found that the white dwarf of VW Hyi was spinning at $ v \sin{i} \sim 600$ km s$^{-1}$, although they subsequently show that even this rotation rate cannot account for the low $L_{X}/L_{opt+UV}$ ratio for this system. Further, Cheng et al. (1997) found that the white dwarf in WZ Sge is rotating at a $v \sin{i} \sim 1200$ km s$^{-1}$. Observations of U Gem show that the white dwarf in that system is rotating at $v \sin{i} \sim 100$ km s$^{-1}$ (Sion et al. 1994), while Mauche et al. (1997) calculate a $v \sin{i} \sim 300$ km s$^{-1}$ for the white dwarf in SS Cyg. Of these systems, U Gem possesses the longest orbital period ($P_{orb} = 4.25$ h) and slowest white dwarf rotation, while WZ Sge has the shortest orbital period ($P_{orb} = 1.36$ h) and the fastest-spinning white dwarf. OY Car, Z Cha, and HT Cas all have orbital periods closer to that of WZ Sge than that of U Gem. On this evidence, it could be that the white dwarf in OY Car is spinning relatively fast, and that this may have the effect of reducing the X-ray flux. 

Alternatively, there may simply be a very low rate of mass transfer through the boundary layer onto the white dwarf. While this may be true for the two observations of HT Cas (it is observed to have occasional low states), it does not explain why the quiescent X-ray luminosities of both OY Car and Z Cha are an order of magnitude lower than typical values.

The real difficulty lies in constraining the value of the column density, $n_{H}$, and it is on this that the luminosity of the system is most dependent. It is perhaps premature to draw conclusions when this factor is such a relative unknown. Figure~\ref{fig:quiespec} shows that a single component column density of $10^{22}$ cm$^{-2}$ is not compatible with the data. Dual-absorption models featuring both neutral and ionized components fare no better because the relative contribution of each component is ill-constrained by the low quality spectrum. The uncertainty in the value of the local absorption in OY Car may well raise some interesting implications, especially for the `iron curtain' model of Horne et al. (1994). In that paper, {\it HST} observations of the quiescent system were decomposed into the individual contributions from the white dwarf, accretion disc, and bright spot. Horne et al. found that a fit to the white dwarf spectrum was only achieved when fitted not only with a white dwarf at $T_{WD} = 16.5 \times 10^{3}$ K, but also with a veiling solar-abundance LTE gas of $T \simeq 10^{4}$ K and $n_{H} \sim 10^{22}$ cm$^{-2}$. This veiling gas was found to have a velocity dispersion of $\Delta v \simeq 60$ km s$^{-1}$, suggesting a physical position at the edge of the disc. The analysis of the `iron curtain' phenomenon has since been extended to Z Cha (Wade, Cheng \& Hubeny 1994). It may be that the `iron curtain' is not a permanent feature, and our observation happened to view the source at an epoch in which the veiling gas was absent. Both our observation and that of Horne et al. (1994) were obtained in the middle of the respective quiescent periods, which suggests that the source of the veiling gas is not directly connected to the outburst mechanism. It may also be that the `iron curtain' is not azimuthally homogeneous, and that the observations obtained by Horne et al. (1994) happened to occur at an orbital phase where the azimuthal distribution of the local absorption was strong.


\section{Conclusions}

We have used {\it ROSAT\/} observations of OY Car to confirm that there is no eclipse of the X-ray flux during superoutburst. Eclipse maps of contemporaneous optical light curves show that three days into the superoutburst decline, the disc was flared with an opening angle of $\alpha \simeq 10^{\circ}$. Taken together with the multiwavelength observations of the 1985 May superoutburst of OY Car in Naylor et al. (1987, 1988), we can confirm that the disc becomes thick enough to obscure the central regions of the disc and boundary layer when in this state. While we now know that discs become physically thick in outburst, we do not yet know of a simple, plausible mechanism that can hold it in place for these considerable lengths of time.

We have difficulty reconciling the measurements of the absorption to the X-ray source in quiescence as derived from measurements in the X-ray and UV. Single-component absorption models with a column density of $10^{22}$ cm$^{-2}$, derived by Horne et al. (1994) from {\it HST} observations, are not compatible with the data; dual-component models do not allow any constraints to be placed on the relative contribution of each component. A possible explanation, that the `iron curtain' is variable, is one that is difficult to test observationally, in that it requires data acquired simultaneously in the UV and X-ray. 

Future work includes further observations of OY Car in quiescence in order to tie down the spectrum of the X-ray source, its position, and the absorption to it. Pratt et al. 1999 uses further {\it ROSAT\/} HRI observations of OY Car in quiescence to constrain the source of the X-ray emission, but detailed modelling of the spectrum and the absorption to the X-ray source requires spectrally resolved data such as obtainable from {\it ASCA}. Looking further into the future, systems such as these will be ideal targets for {\it XMM}, and will make excellent use of the simultaneous optical monitoring.


\section*{Acknowledgements}

We thank K. Beuermann, F.V. Hessmann, B. G\"{a}nsicke and K. Reinsch for useful discussions. We acknowledge the data analysis facilities provided by the Starlink Project, which is run by CCLRC on behalf of PPARC. GWP is in receipt of a University of Central Lancashire research studentship.



\end{document}